\documentclass[12pt]{article}

\usepackage{scicite}

\usepackage{xcolor}
\usepackage{times}
\usepackage{changes}
\usepackage{graphicx}
\usepackage{amssymb}
\usepackage{epstopdf}
\usepackage{amsmath}
\usepackage{color}
\usepackage{hyperref}
\usepackage{bbold}
\usepackage{float}
\usepackage{physics}

\newenvironment{sciabstract}{%
\begin{quote} \bf}
{\end{quote}}

\title{Confinement and Entanglement Dynamics \\on a Digital Quantum Computer} 

\author
{Joseph Vovrosh$^{1\ast}$, Johannes Knolle$^{2,3,1}$\\
\\
\normalsize{$^{1}$Blackett Laboratory, Imperial College London,}\\
\normalsize{London SW7 2AZ, United Kingdom}\\
\normalsize{$^{2}$Department of Physics TQM, Technische Universit{\"a}t M{\"u}nchen,}\\
\normalsize{James-Franck-Stra{\ss}e 1, D-85748 Garching, Germany}\\
\normalsize{$^{3}$Munich Center for Quantum Science and Technology (MCQST),}\\
\normalsize{80799 Munich, Germany}\\
\\
\normalsize{$^\ast$To whom correspondence should be addressed; E-mail:  jwv14@ic.ac.uk.}
}

\date{}

\begin{document} 

\baselineskip24pt

\maketitle 

\begin{sciabstract}
Confinement describes the phenomenon when the attraction between two particles grows with their distance, most prominently found in quantum chromodynamics (QCD) between quarks.  In condensed matter physics, confinement can appear in quantum spin chains, for example, in the one dimensional transverse field Ising model (TFIM) with an additional longitudinal field~\cite{mccoy1978two,fonseca2003ising}, famously observed in the quantum material cobalt niobate~\cite{experiment, lake2010confinement} or in optical lattices~\cite{simon2011quantum}. Here, we establish that state-of-the-art quantum computers have reached  quantum simulation capabilities to explore confinement physics in spin chains. We report quantitative confinement signatures of the TFIM on an IBM quantum computer observed via two distinct velocities for information propagation from domain walls and their mesonic bound states. We also find the confinement induced slow down of entanglement spreading~\cite{nature} by implementing randomized measurement protocols for the second order R\'enyi entanglement entropy~\cite{brydges2019probing}. Our results are a crucial step for probing non-perturbative interacting quantum phenomena on digital quantum computers beyond the capabilities of classical hardware. 
\end{sciabstract}

\paragraph*{Introduction.---}  Quantum computers are proposed to out-perform their classical counterparts for selected applications~\cite{nielsen2002quantum}. It is Richard Feynman's prediction from 1982 that a quantum device would have the ability to directly simulate quantum systems which has most potential for solving a number of long-standing fundamental problems in science~\cite{feynman1982simulating,lloyd1996universal,georgescu2014quantum}, for example in chemistry~\cite{kandala2017hardware} or for lattice gauge theories (LGT) relevant in high energy physics~\cite{martinez2016real,jordan2012quantum,zohar2015quantum}. 

In recent years there has been a tremendous push in order to realise a digital quantum computer, e.g., based on superconducting circuits. Despite these efforts, current working computers are described as Noisy Intermediate-Scale Quantum (NISQ) devices~\cite{preskill2018quantum}, which do not have enough qubits or small enough errors to perform error correction. The uses of NISQ devices are still in question but we show here that they have reached capabilities for simulating quantum confinement physics.  

Our basic understanding of confinement in QCD is limited because it is an example of a non-perturbative quantum many body effect. LGT descriptions thereof~\cite{brambilla2014qcd} are hard to simulate on classical computers and remain beyond the reach of current NISQ devices. As a first step for proving the usefulness of quantum computers as quantum simulators one can study one dimensional lattice systems of condensed matter physics displaying similar confinement physics. Examples include, the transverse field Ising model (TFIM) with long range interactions \cite{long}, the lattice Schwinger model~\cite{martinez2016real}, the XXZ spin-1/2 chain \cite{microscopic} and, the  model considered here, the TFIM with an additional longitudinal field~\cite{nature,James2019nonthermal}. The pure TFIM has free fermion excitations that correspond to domain walls between spin-aligned segments. An additional longitudinal field gives rise to an emergent confining potential between these fermionic excitations resulting in `mesonic' bound states. Fig. 1 (a+b) shows the different velocities of free (dashed) and bound (solid) particles that govern the time evolution of correlation spreading. Together with the confinement induced halting of entanglement spreading, Fig. 1 (c), this provides direct signatures of confinement physics on a digital quantum computer. 

The TFIM has been studied with analytical methods~\cite{fonseca2003ising,mussardo2011integrability,rutkevich2008energy} but the full time evolution of the non-integrable model with a longitudinal field has been restricted to numerical simulations for limited system sizes or time windows, e.g. with the density matrix renormalization group (DMRG)~\cite{nature}. In general, out-of-equilibrium quantum dynamics of many-body systems are notoriously difficult to simulate with classical computers because the memory required scales exponentially with system size. In principle, quantum computers are free of such problems, however, available NISQ devices come with their own limitations. Firstly, they only have a restricted number of available qubits. Secondly, their large errors when executing a quantum circuit limit circuit depth and in turn the accessible simulation time. Nevertheless, there have already been promising results for the magnetization dynamics of different spin chains~\cite{cervera2018exact, zhukov2018algorithmic,francis2019quantum}. However, up to now the accuracy of the devices was barely enough to qualitatively distinguish genuine interaction from disorder/noise effects~\cite{smith2019simulating}. Here, we take the next step and report digital quantum simulation of confinement in out-of-equilibrium dynamics of spin chains of up to nine spins on the latest IBM machines code-named Boeblingen and Paris.

\paragraph*{Model and the two kink subspace.---} The one dimensional TFIM with an additional longitudinal field is described by the following Hamiltonian
\begin{equation}
\label{Eq1}
    H = -J\bigg[\sum_{i=0}^{L-2}\sigma_i^{z}\sigma_{i+1}^{z} + h_x\sum_{i=0}^{L-1} \sigma_i^{x}+ h_z\sum_{i} \sigma_i^{z}\bigg],
\end{equation}
where $\sigma_i^{\alpha}$, $\alpha \in {x\text{, }y\text{, }z}$ are the Pauli matrices acting on the $i^{th}$ site, $i \in \{0,1,2,...,L-1\}$, $L$ is the length of the chain, $J$ is the Ising exchange of nearest neighbour spin $1/2$ and $h_{x/z}$ are the relative strengths of the transverse and longitudinal fields. For $h_z = 0$, the TFIM can be exactly diagonalised via Jordan-Wigner transformation and describes free fermions. Here we restrict ourselves to transverse field strengths below its critical value, $h_c = J$, in the ordered phase, where fermions are domain wall (or kink) excitations $\ket{...\uparrow\uparrow\downarrow\downarrow...}$ aligned in the $z$ direction. 
The longitudinal field then gives rise to a confining potential between kinks strongly affecting the non-equilibrium dynamics of the system. An established way to elucidate the confinement physics is to study the dynamics in a restricted two-kink subspace which not only allows us to predict analytically the velocities and masses of the mesons, see Fig.1, but crucially for this work, it also forms the basis of our error mitigation protocol. We project Eq.(1) into the two kink subspace written in the basis
$\ket{j,n} = \ket{\uparrow\uparrow...\uparrow\downarrow_j...\downarrow_{j+n-1}\uparrow..\uparrow\uparrow}$.
This gives $\mathcal{H} = P^{-1}HP$, in which $P$ is the projection operator and (up to constant terms).
\begin{equation}
    \mathcal{H} =\sum_{j,n} \bigg\{-h_x\big[\ket{j,n+1}+\ket{j,n-1}+\ket{j+1,n-1}+\ket{j-1,n+1}\big]\bra{j,n}  +V(n)\ket{j,n}\bra{j,n}\bigg\},
    \label{Eq2}
\end{equation}
with $V(n) = 2h_zn$. The first term of this subspace Hamiltonian is a kinetic term that allows the kinks to `hop', and the second term is the effective potential, $V(n)$, linearly increasing with kink separation $n$. Thus, the out-of-equilibrium motion of kinks will be similar to that of quarks; pairs of kinks that are produced propagate in opposite directions until the confining potential halts their motion and pulls them back, leading to oscillatory motion, this is what we call a meson. These mesonic bound states of kinks are then able to propagate as a pair with a much slower velocity than if they were free.

\begin{figure}[H]
\vspace{-2.5cm}
    \centering
    \includegraphics[width=1\textwidth]{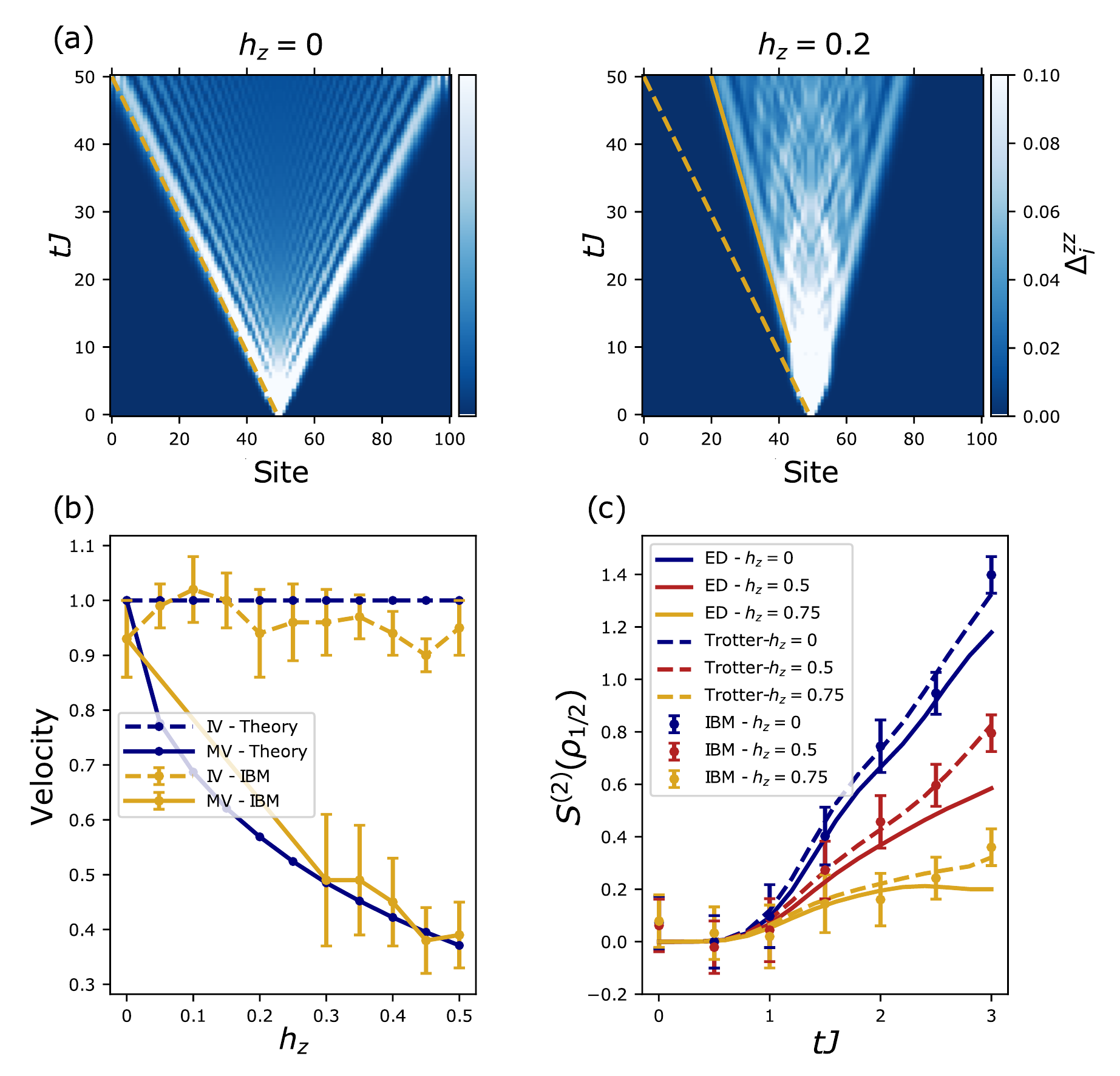}
    \caption{\textbf{Velocities and entanglement Renyi entropy from the IBM device}. (a) Real time dynamics of domain wall positions within the two kink subspace following a quantum quench to the TFIM without and with additional longitudinal field $h_z$. Here, $L=101$, $h_x=0.5$ and the initial state is ferromagnetic with a single flipped spin in the centre. For $h_z=0$ the light cone structure of free particles is visible. For the confining case $h_z\neq0$ two velocities are observable, an initial velocity (dashed) equal to the free case and the meson velocity (solid) at longer times. (b) Comparisons of the two velocities as measured on the IBM quantum computer ($h_x=0.5$ and $L=9$) after error mitigation and as theoretically predicted. Error bars displayed are the standard deviation of a range of velocities obtained, more details are given in the Supplementary Material. (c) Data from randomized measurements for the half chain second order R\'enyi entropy after a global quantum quench to the TFIM with varying a longitudinal field strengths on the state $\ket{\frac{L}{2}-1,2}$ for $h_x=0.5$, $L=6$. The balistic entanglement growth of the free case is suppressed because of confinement for increasing longitudinal field. Here, error bars are calculated by jacknife resampling. The inherent error in the IBM device leads to offset which has been removed, see the Supplementary Material for details.}
    \label{Fig1}
\end{figure}
\newpage

\paragraph*{Signatures of confinement.---} A hallmark signature of confinement is the formation of mesons whose properties, e.g. masses, have been measured with different observables depending on the experimental or numerical feasibility~\cite{experiment, lake2010confinement,simon2011quantum,nature}. In order to use NISQ devices as new tools for quantum simulations it is necessary to carefully design the measurement set-up to obtain unambiguous signatures for the available system sizes and time windows. Among the many different protocols we have checked, the three following measures can give qualitative as well as quantitative results on the IBM devices.

i) Confinement physics can be observed in the probability dynamics of kinks as a function of time given by
\begin{equation}
\label{Eq4}
    \Delta_{i}^{zz}=\bra{\psi(t)}\frac{1}{2}(1-\sigma^{z}_{i}\sigma^{z}_{i+1})\ket{\psi(t)}.
\end{equation}
This function gives the probability distribution of kinks along a chain. Thus, it provides a very clear picture of kink motion. In fact, it not only shows the kinks position with time, it also shows the
mesons position once they form. We have benchmarked that both of the two velocities can be extracted from $\Delta^{zz}_i$ with quantitative agreement to the theory described above.

ii) Another key signature is the suppression of half chain entanglement entropy spreading due to confinement. In the free case, $h_z=0$, the entanglement entropy is expected to increase linearly~\cite{fagotti2008evolution}. However, with a non-zero confining field $h_z$ this growth is suppressed in a characteristic fashion~\cite{nature}.
In general, entanglement entropy is not easy to calculate on a real quantum computer as it requires some form of state tomography. Here, building on recent progress for randomized measurement algorithms~\cite{brydges2019probing} we are able to measure -- for the first time on a digital quantum computer -- the second order R\'eyni entanglement defined as

\begin{equation}
    S^{(2)}(\rho_A) = - \log_2\Tr(\rho_A^2).
\end{equation}

Here, $\rho_A$ is the reduced density matrix for the half-chain subsystem A. With a repeated measurement of a set of random single qubit gates on each site in the subsystem A, $S^{(2)}(\rho_A)$ can be approximated by 

\begin{equation}
    S^{(2)}(\rho_A) = - \log_2\bar{X},\text {with }X = 2^{N_A}\sum_{s_A,s'_A}-2^{D[s_A,s'_A]}P(s_A)P(s'_A),
\end{equation}

\noindent where $s_A$ denotes a measurement outcome, $P(s_A)$ is the probability of measuring $s_A$, $D[s_A,s'_A]$ is the Hamming distance between $s_A$ and $s'_A$, $N_A$ is the dimension of $A$ and $\bar{X}$ denotes the ensemble average of $X$ over the set of different random single qubit gates used~\cite{brydges2019probing}.

iii) The last viable diagnostic of confinement are the probability maps of kink positions~\cite{microscopic}. After time evolution the probabilities of the first kink position with respect to the position of the second kink is mapped to show that, in the presence of an additional longitudinal field, it is favourable for the two kinks to reside close to each other, i.e. the kinks form a meson.

\paragraph*{Error mitigation and post selection.---} When implementing dynamics on the IBM quantum computers there are four main sources of error: initialisation and measurement error, single qubit gate error, controlled-NOT (CNOT) gate error and decoherence. A crucial ingredient for obtaining quantitative results are the following error mitigations.

i) The best subset of qubits is chosen. This is done by calculating the average error for each subset of qubits with the desired topology, a chain of length $L$, within the machine. Error types are not weighted evenly as the gate error is more important than the readout error for the protocols used in this work. Although this method is not scalable with increasing number of qubits, it is well suited for current devices.

ii) The initial states considered here have an inherent inversion symmetry around the centre site such that the data should reflect this symmetry. However, because of inhomogeneous errors in the quantum computer there are sizeable deviations which we correct by averaging the data and with its mirror image.

iii) The last and crucial error mitigation technique is the projection of the data into the two kink subspace. For our initial states and quench set-up, the error-free time evolution mainly takes place within this subspace, and crucially, it contains the desired confinement physics. Hence, this post selection to the two kink subspace is a viable tool for eliminating errors, more details are given in the Supplementary Material.  

With the error mitigation described above, as well as repeating results on different days~\cite{smith2019simulating}, simulations for meson velocities and probability maps with up to nine qubits and times of up to $tJ=8$ were obtained. For the second order R\'enyi entanglement entropy it is necessary to apply random unitaries before measurements such that only mitigation technique i) can be used. Times that can be simulated for the second order R\'enyi entropy are up to $tJ\sim3$.

\begin{figure}
    \centering
    \includegraphics[width=1\textwidth]{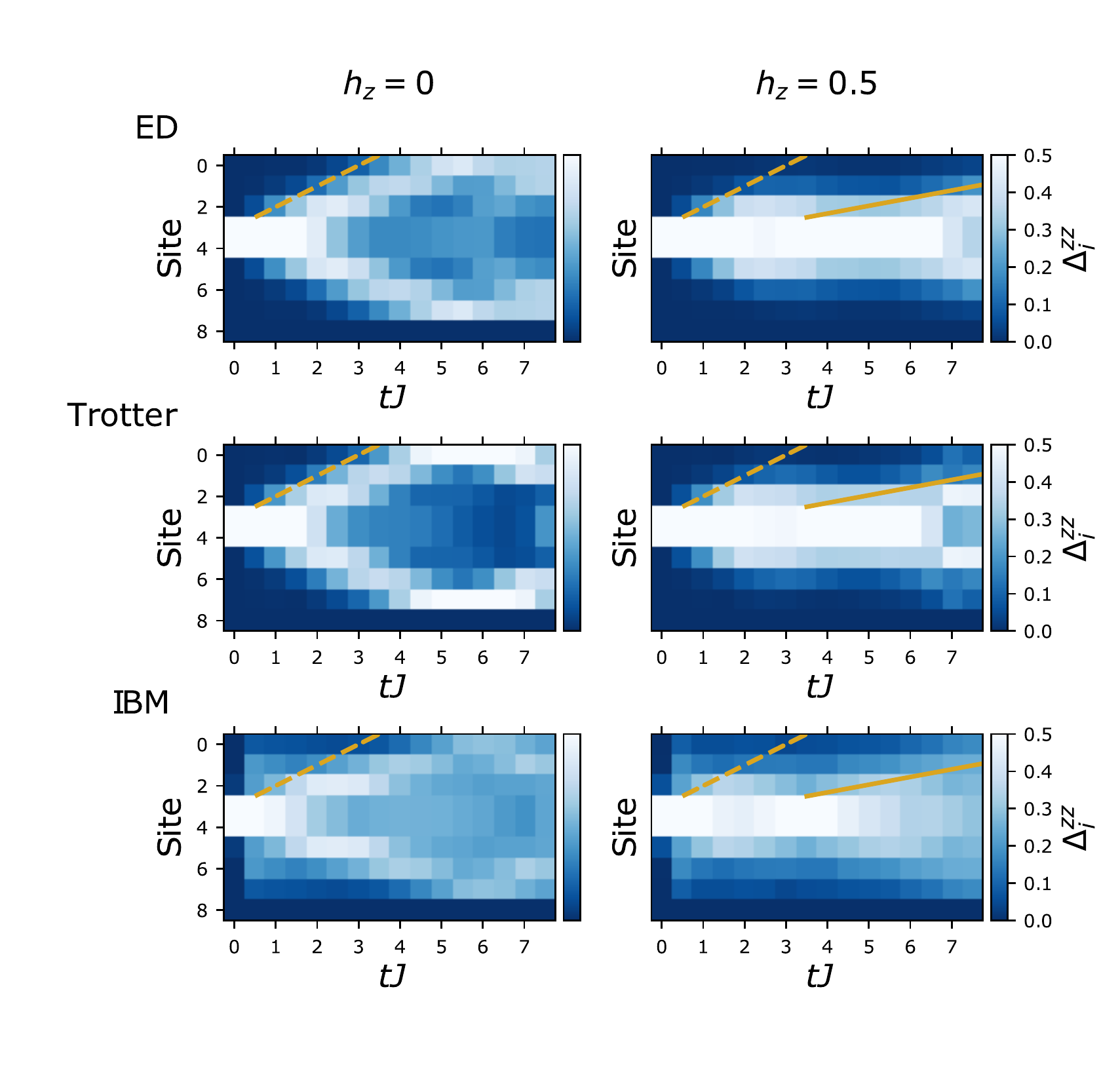}
    \caption{\textbf{Time evolution of probability dynamics of kinks}. Data for $\Delta_{i}^{zz}$ after a global quantum quench to the TFIM with and without a longitudinal field starting from the state $\ket{\frac{L-1}{2},1}$. In all presented data $h_x=0.5$ and $L=9$. The graphs on the left show the free kink case, $h_z=0$ and the graphs on the right the confined one $h_z =0.5$. Clear suppression of the kink separation can be seen in the latter as well as the emergence of a second slower velocity -- both signatures of confinement.}
    \label{Fig3}
\end{figure}

\paragraph*{Results.---} Data for the probability maps as well as $\Delta^{zz}_i$ was collected from the IBM computers Boeblingen which has a total of 20 qubits. We employed a global quench protocol from an initially aligned state with a single spin flipped at the centre (the state $\ket{\frac{L-1}{2},1}$) to the TFIM with and without a longitudinal field. Randomized measurements for the second order R\'enyi entanglement entropy were carried out on the IBM device Paris with a total of 27 qubits. We employed a global quench from an initially aligned state with two spins flipped at the centre (the state $\ket{\frac{L}{2}-1,2}$). We used open boundary conditions in all cases.

i) In Fig.\ref{Fig3} we show results for $\Delta_i^{zz}$ (with $h_z=0, 0.5$ and $h_x=0.5$) from the IBM machine compared to continuous time exact diagonalization (ED) and trotterised ED, both projected to the two kink subspace. The short time dynamics is governed by the free motion of kinks (dashed) before the bound states form and propagate at the slower meson velocity (solid). From these results, initial velocities and second velocities were extracted for varying longitudinal fields and compared to theoretically predicted values, which are summarised in Fig.\ref{Fig1}(b). Details how the velocities and error bars were obtained are given in the Supplementary Material. The extracted meson velocities, shown in Fig.\ref{Fig1}(b), match quantitatively the  predictions from the two kink subspace analysis.

ii) The second order R\'enyi entanglement entropy results are presented in Fig.\ref{Fig1}(c). Here, we compare the exact results calculated via ED and trotterisation with the data from the IBM device. It reproduces the suppression of entanglement spreading that depends on the strength of the confining longitudinal field. We note that the inherent error on IBM devices leads to a constant shift which is removed in Fig.\ref{Fig1}(c), more details are given in the Supplementary Material.

iii) Finally, Fig.\ref{Fig5}(a) displays the probability maps of kink motion collected from the quantum computer. These maps show how the longitudinal field favours the two kinks to stay together as expected by confinement dynamics~\cite{microscopic}.

To corroborate our findings, it is crucial to confirm that the halting of domain wall spreading for increasing $h_z$ arises from coherent quantum dynamics and not just disorder or noise from the machine which have plagued previous attempts~\cite{smith2019simulating}. In Fig.\ref{Fig5}(b) we show the evolution of the local magnetisation for a quench with $h_x=h_z=0.5$ and $L=7$. Clear oscillatory patterns of the confined kink motion are observed which provide direct evidence of higher order interaction effects and not a simple featureless decay of correlations. 

 \begin{figure}[H]
 \vspace{-2cm}
    \centering
    \includegraphics[width=1\textwidth]{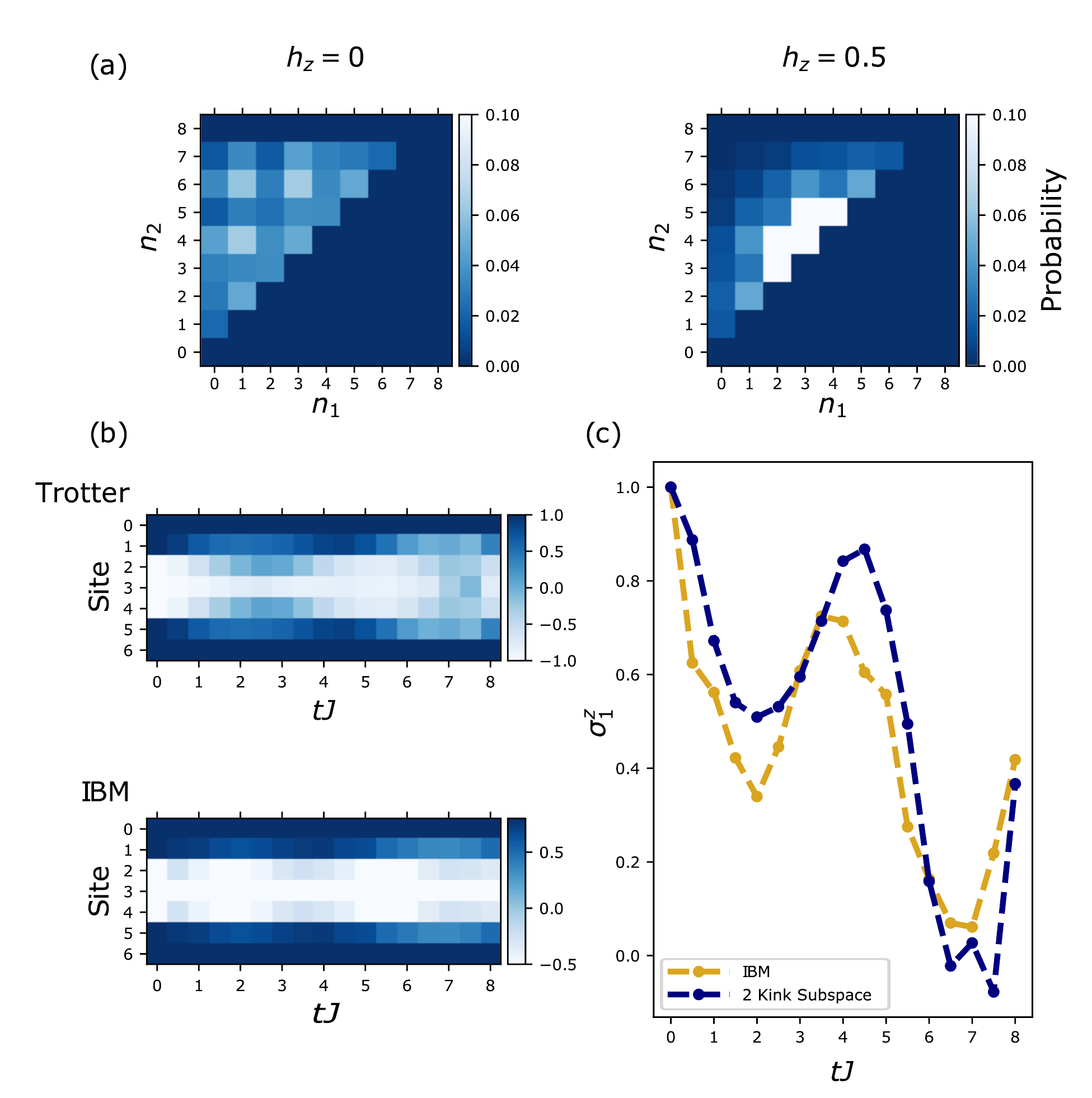}
    \caption{\textbf{Probability maps and time evolution of local magnetisation}. (a) Data from the IBM device of the probability maps of kink position after a global quantum quench with and without a longitudinal field (for $h_x=0.5$ and $L=9$) on the state $\ket{\frac{L-1}{2},1}$. The graph on the left shows the free kink case, $h_z=0$ and in the graph on the right $h_z =0.5$. Clearly if $h_z=0$, the the kinks have no preference to remain close together as there is no confining field. However, for $h_z = 0.5$, the kinks have a much larger probability to reside close to one another than being separated. (b) The local magnetisation after a global quantum quench with $h_x=h_z=0.5$ and $L=7$ on the state $\ket{\frac{L-1}{2}-1,3}$. These results show clear oscillatory motion of kinks. This is a high order effect that is only seen with interactions and not just disorder. (c) The local magnetisation of the first qubit before symmetrisation, $\sigma_1^z$, is shown explicitly, highlighting the oscillatory behaviour captured by the quantum computer.}
    \label{Fig5}
\end{figure}
\newpage
\paragraph*{Discussion.---}
We have established that current state-of-the-art quantum computers are able to simulate non-perturbative quantum effects like confinement. Using a specially designed quench set-up has allowed us to show confinement signatures and the formation of domain wall bound states in the paradigmatic TFIM with a longitudinal field. Randomized measurement protocols have enabled us to show the confinement-induced slow-down of entanglement spreading on a digital quantum computer. Next on the agenda will be quantum simulations of confinement effects in spin chains as discussed in relation with scattering experiments of real materials~\cite{experiment,lake2010confinement,wang2018experimental}. On a different front, digital quantum simulations and especially entanglement measurements as presented here will help to further our understanding of non-ergodic quantum dynamics~\cite{Banuls2011strong}, for example the interplay of confinement and quantum many body scars~\cite{James2019nonthermal,van2020quantum}. 

Our benchmark results are a crucial step towards the simulation of many-body quantum phenomena beyond the reach of classical computers. The added advantage to similar quantum simulation endeavours in cold atomic gases~\cite{simon2011quantum} or trapped ion quantum simulators \cite{becker2020observation} is the ease of initial/final-state preparation/selection, as well as the potential freedom to engineer more complicated theories also in higher dimensions, e.g., for quantum field/gauge theories~\cite{jordan2012quantum,zohar2015quantum}. For example, a next step for digital quantum simulations would be to simulate the Schwinger model in order to observe out-of-equilibrium properties of $1 + 1$ dimensional quantum electrodynamics~\cite{martinez2016real} before going to higher dimensions. The good news for NISQ limited devices is that signatures of confinement, like pair production or string breaking~\cite{magnifico2019real}, are visible already at short times for moderate system sizes. In the long run, a potentially disruptive advantage of digital quantum simulators is the ease of access to experimental hardware. We provide a first example how to use remote access to a NISQ device as a new numerical tool which can perform specific experiments, for example on confinement dynamics, without the need for purpose built set ups \cite{becker2020observation}.

This work highlights the capabilities of quantum computers in the NISQ era. They can already deliver on Feynman’s original quantum simulation promise --- for the time being at least for phenomena like confinement which are observable in intermediate-time dynamics and for moderate system sizes.  

\bibliography{scibib}

\begin{thebibliography}{10}

\bibitem{mccoy1978two}
B.~M. McCoy, T.~T. Wu, {\it Phys. Rev. D\/} {\bf 18}, 1259 (1978).

\bibitem{fonseca2003ising}
P.~Fonseca, A.~Zamolodchikov, {\it Journal of statistical physics\/} {\bf 110},
  527 (2003).

\bibitem{experiment}
R.~Coldea, {\it et~al.\/}, {\it Science\/} {\bf 327}, 177 (2010).

\bibitem{lake2010confinement}
B.~Lake, {\it et~al.\/}, {\it Nature Physics\/} {\bf 6}, 50 (2010).

\bibitem{simon2011quantum}
J.~Simon, {\it et~al.\/}, {\it Nature\/} {\bf 472}, 307 (2011).

\bibitem{nature}
M.~Kormos, M.~Collura, G.~Tak{\'a}cs, P.~Calabrese, {\it Nature Physics\/} {\bf
  13}, 246 (2017).

\bibitem{brydges2019probing}
T.~Brydges, {\it et~al.\/}, {\it Science\/} {\bf 364}, 260 (2019).

\bibitem{nielsen2002quantum}
M.~A. Nielsen, I.~Chuang, Quantum computation and quantum information (2002).

\bibitem{feynman1982simulating}
R.~P. Feynman, {\it International journal of theoretical physics\/} {\bf 21},
  467 (1982).

\bibitem{lloyd1996universal}
S.~Lloyd, {\it Science\/} pp. 1073--1078 (1996).

\bibitem{georgescu2014quantum}
I.~M. Georgescu, S.~Ashhab, F.~Nori, {\it Reviews of Modern Physics\/} {\bf
  86}, 153 (2014).

\bibitem{kandala2017hardware}
A.~Kandala, {\it et~al.\/}, {\it Nature\/} {\bf 549}, 242 (2017).

\bibitem{martinez2016real}
E.~A. Martinez, {\it et~al.\/}, {\it Nature\/} {\bf 534}, 516 (2016).

\bibitem{jordan2012quantum}
S.~P. Jordan, K.~S. Lee, J.~Preskill, {\it Science\/} {\bf 336}, 1130 (2012).

\bibitem{zohar2015quantum}
E.~Zohar, J.~I. Cirac, B.~Reznik, {\it Reports on Progress in Physics\/} {\bf
  79}, 014401 (2015).

\bibitem{preskill2018quantum}
J.~Preskill, {\it Quantum\/} {\bf 2}, 79 (2018).

\bibitem{brambilla2014qcd}
N.~Brambilla, {\it et~al.\/}, {\it The European Physical Journal C\/} {\bf 74},
  2981 (2014).

\bibitem{long}
F.~Liu, {\it et~al.\/}, {\it arXiv preprint arXiv:1810.02365\/}  (2018).

\bibitem{microscopic}
T.~Fukuhara, {\it et~al.\/}, {\it Nature\/} {\bf 502}, 76 (2013).

\bibitem{James2019nonthermal}
A.~J.~A. James, R.~M. Konik, N.~J. Robinson, {\it Phys. Rev. Lett.\/} {\bf
  122}, 130603 (2019).

\bibitem{mussardo2011integrability}
G.~Mussardo, {\it Journal of Statistical Mechanics: Theory and Experiment\/}
  {\bf 2011}, P01002 (2011).

\bibitem{rutkevich2008energy}
S.~Rutkevich, {\it Journal of Statistical Physics\/} {\bf 131}, 917 (2008).

\bibitem{cervera2018exact}
A.~Cervera-Lierta, {\it Quantum\/} {\bf 2}, 114 (2018).

\bibitem{zhukov2018algorithmic}
A.~Zhukov, S.~Remizov, W.~Pogosov, Y.~E. Lozovik, {\it Quantum Information
  Processing\/} {\bf 17}, 223 (2018).

\bibitem{francis2019quantum}
A.~Francis, J.~Freericks, A.~Kemper, {\it arXiv preprint arXiv:1909.05701\/}
  (2019).

\bibitem{smith2019simulating}
A.~Smith, M.~Kim, F.~Pollmann, J.~Knolle, {\it npj Quantum Information\/} {\bf
  5} (2019).

\bibitem{fagotti2008evolution}
M.~Fagotti, P.~Calabrese, {\it Physical Review A\/} {\bf 78}, 010306 (2008).

\bibitem{wang2018experimental}
Z.~Wang, {\it et~al.\/}, {\it Nature\/} {\bf 554}, 219 (2018).

\bibitem{Banuls2011strong}
M.~C. Ba\~nuls, J.~I. Cirac, M.~B. Hastings, {\it Phys. Rev. Lett.\/} {\bf
  106}, 050405 (2011).

\bibitem{van2020quantum}
B.~van Voorden, J.~Min{\'a}{\v{r}}, K.~Schoutens, {\it arXiv preprint
  arXiv:2003.13597\/}  (2020).

\bibitem{becker2020observation}
P.~Becker, {\it et~al.\/}, {\it Bulletin of the American Physical Society\/}
  (2020).

\bibitem{magnifico2019real}
G.~Magnifico, {\it et~al.\/}, {\it arXiv preprint arXiv:1909.04821\/}  (2019).

\end{thebibliography}

\bibliographystyle{Science}

\paragraph*{Acknowledgments.---} We are grateful for discussions with Hongzheng Zhao, Adam Smith, Kiran Kholsa, Markus Heyl, Frank Pollmann and Myungshik Kim. We particularly thank Adam Smith for help with the initial qiskit implementation and comments on the manuscript and Kiran Kholsa for help with implementation to measure R\'enyi entanglement entropies. We acknowledge the Samsung Advanced Institute of Technology Global Research Partnership and travel support via the Imperial-TUM flagship partnership.

\paragraph*{Data availability.---} All data is available upon reasonable request.

\newpage
\section*{Methods and Supplementary Material}

\paragraph*{Calculation of the meson velocities.---} In order to calculate the meson velocities we proceed by taking the Fourier transform of Eq.(2) over $j$, $\ket{k,n}=\frac{1}{\sqrt{L}}\sum_j\exp(-ik\frac{n}{2}-ikj)\ket{j,n}$, obtaining
\begin{multline}
    \mathcal{H} = \sum_{k,n} \big[V(n)\ket{k,n}\bra{k,n} +2h_x\cos{\frac{k}{2}}\big(\ket{k,n}\bra{k,n-1}+\ket{k,n}\bra{k,n+1}\big)\big]
\end{multline}
which can be diagonalised using the transformation
\begin{equation}
    \ket{k,\alpha} = \sum_{n}C_\alpha\mathcal{J}_{n-\nu_{k,\alpha}}(x_k)\ket{k,n}.
\end{equation}
in which $\nu_{k,\alpha} = \frac{E_{k,\alpha}}{2h_x}$, $x_k = \frac{2h_z\cos\frac{k}{2}}{h_x}$, $\mathcal{J}$ is the Bessel function of the first kind and the coefficient $C_\alpha$ is used for normalisation. The energy levels $E_{k,\alpha}$ can be computed via the boundary condition that $\mathcal{J}_{-\nu_{k,\alpha}}(x_k) = 0$. Continuing with the analogy from QCD, the masses of the mesons formed by the domain wall pairs can be calculated via the difference between energy levels and the ground state. As well as these, the allowed velocities of the mesons can be calculated as the maximal gradient of each energy level, see Fig. 1(b). The observation of these quantities then provides direct evidence of the underlying confinement dynamics. For example, in Fig 1 (a), with $h_z=0$ one velocity is seen and corresponds to the free kink motion described by the TFIM. However, if $h_z=0.2$ there are two velocities. The initial velocity from $t=0$ until $t\sim4J$ again shows free kink motion, while for times $t>15J$ the slower velocity of the meson governs the dynamics.

\paragraph*{Implementation on the IBM quantum computer.---}
In order to implement quench dynamics onto a quantum computer one must decompose the time evolution operator, $U(t) = e^{-iHt}$ into one and two qubit gates. To accomplish this, the standard Suzki-Trotter decomposition, commonly known as trotterisation, is used. Trotterisation discretises time based on the fact that for two non-commuting operators $A$ and $B$,   
$e^{A+B} = \lim_{n\rightarrow\infty}\big(e^\frac{A}{n}e^\frac{B}{n}\big)^n$.
Using this for a Hamiltonian of the form $H = A+B$, the time evolution operator can be written as
\begin{equation}
\label{Eq9}
    U( t)=e^{-iHt}=e^{-iAt}e^{-iBt}+O(t^2).
\end{equation}

Clearly, the Suzki-Trotter decomposition will only give reliable results for small times. Thus, in order to simulate long time periods on must use multiple trotter steps, $U(t) \sim U(\Delta t)$ $U(\Delta t)$ $U(\Delta t)$..., here $U(\Delta t)$ is given by the approximation in Eq.(8) The number of trotter steps required depends on the length of time that is being simulated. There are extensions to this approximation to further reduce the resulting error. Building on our previous work, we use the symmetric decomposition~\cite{smith2019simulating} given by:
\begin{equation}
    U( t)=e^{-iHt}=e^{-iA\frac{t}{2}}e^{-iBt}e^{-iCt}e^{-iA\frac{t}{2}}+O(t^3),
\end{equation}
with $A = -Jh_x\sum_{i=1}^L \sigma_i^{x}$, $B =  -Jh_z\sum_{i=1}^L \sigma_i^{z}$ and $C = -J\sum_{i}^L\sigma_i^{z}\sigma_{i+1}^{z}$. Note that $e^{-iBt}$ is not symmetrised in this expression as $[B,C]=0$. A schematic of the gate sequence required to implement this is given in Fig. 2.

\begin{figure}[H]
    \centering
    \includegraphics[width=0.5 \textwidth]{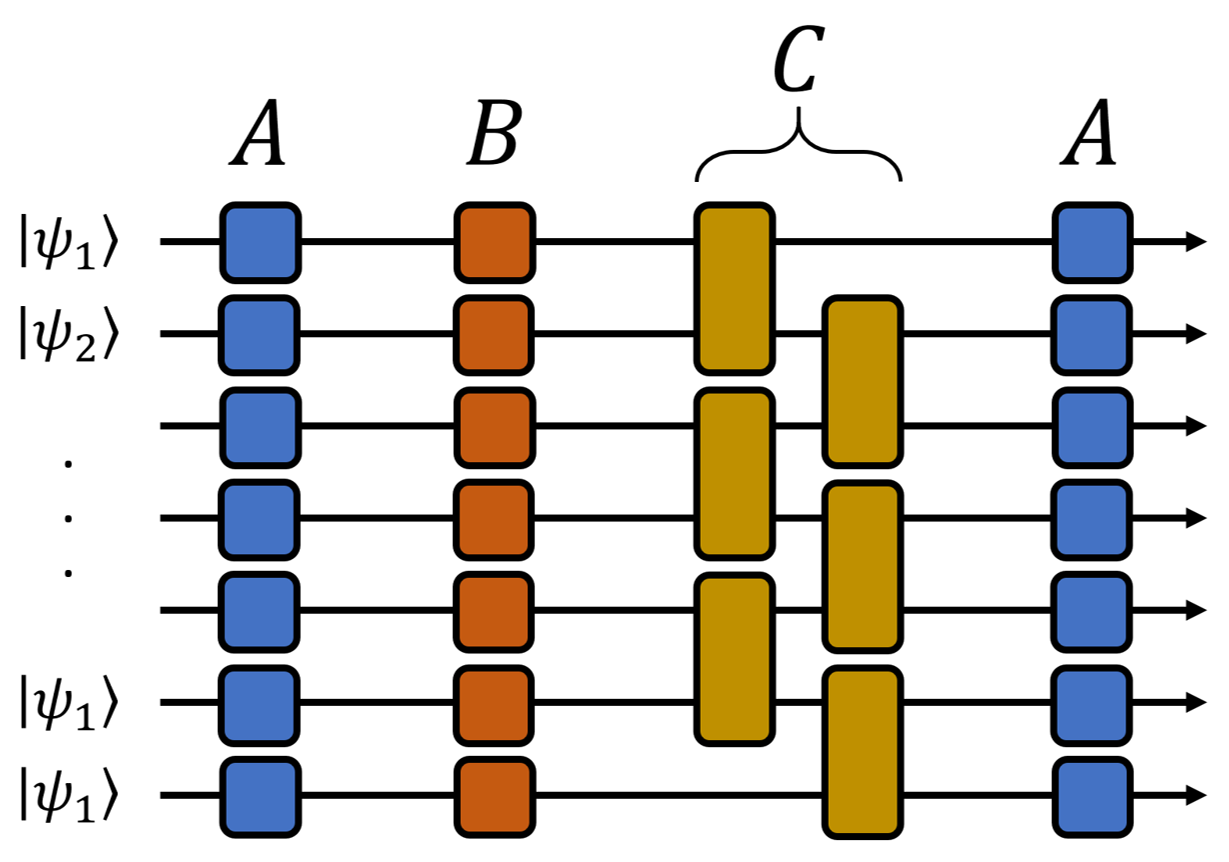}
    \caption{\textbf{The gate sequence in each trotter step performed on the quantum computer}. Here, $A = e^{\frac{itJh_x}{2}\sum_{i=1}^L \sigma_i^{x}}$, $B =  e^{itJh_z\sum_{i=1}^L \sigma_i^{z}}$ and $C = e^{itJ\sum_{i}^L\sigma_i^{z}\sigma_{i+1}^{z}}$.}
    \label{Fig2}
\end{figure}

\newpage
\paragraph*{Velocity extraction.---} In order to extract the initial free kink and later meson velocities from the mitigated data obtained by the IBM device the gradient of the light cone formed in $\Delta_i^{zz}$ data is computed. Due to the inherent error in the NISQ device, there is a level of uncertainty in the saturation levels that should be used. Thus, a range of velocities are computed and the averages and standard deviations are presented in Fig.\ref{Fig1}.

\begin{figure}[H]
    \centering
    \includegraphics[width=1 \textwidth]{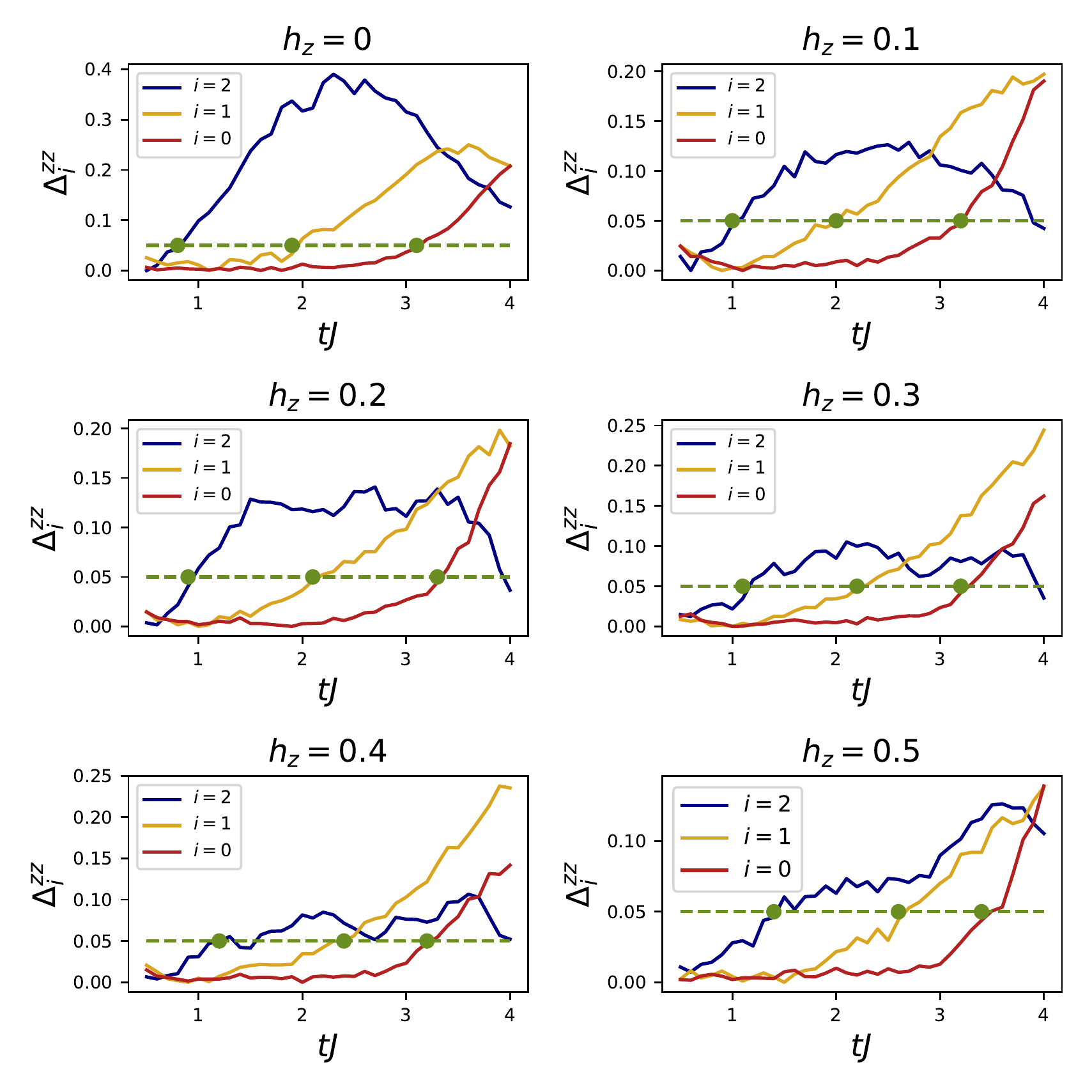}
    \caption{\textbf{$\Delta_i^{zz}$ data used to calculated initial velocities of kinks}. Re-scaled data for $\Delta_{i}^{zz}$ after a global quantum quench to the TFIM with a longitudinal field on the state $\ket{\frac{L-1}{2},1}$. In all presented data $h_x=0.5$ and $L=9$. Green points correspond to the $(x,t)$ coordinates used to calculate the initial velocities for a saturation level of $S = 0.05$. Each figure shows clearly the free kink movement from sites $2$ to $0$. The velocities measured depend heavily on the saturation levels used hence the use of a large range to get an average.}
    \label{Fig6}
\end{figure}

To attain the initial velocities, data for quench dynamics up to $t=4J$ was simulated using four trotter steps. In these results there are initialisation errors that would cover up the free kink velocity unless they are removed. Fig.\ref{Fig6} shows $\Delta_i^{zz}$ data collected for sites $i \in \{0,1,2\}$ after forcing the minimum of each to be zero, this removes the effect of the initialisation error. To calculate a velocity one can choose a saturation level, $S$, and find the times at which $\Delta_i^{zz}$ surpasses this number for each site $i \in \{0,1,2\}$. This gives three $(x,t)$ coordinates for which the gradient of a line of best fit gives the velocity. When computing the light cone gradients for initial velocities the same range of saturation levels were used for each value of $h_z$ for consistency.

\begin{figure}[H]
    \centering
    \includegraphics[width=1 \textwidth]{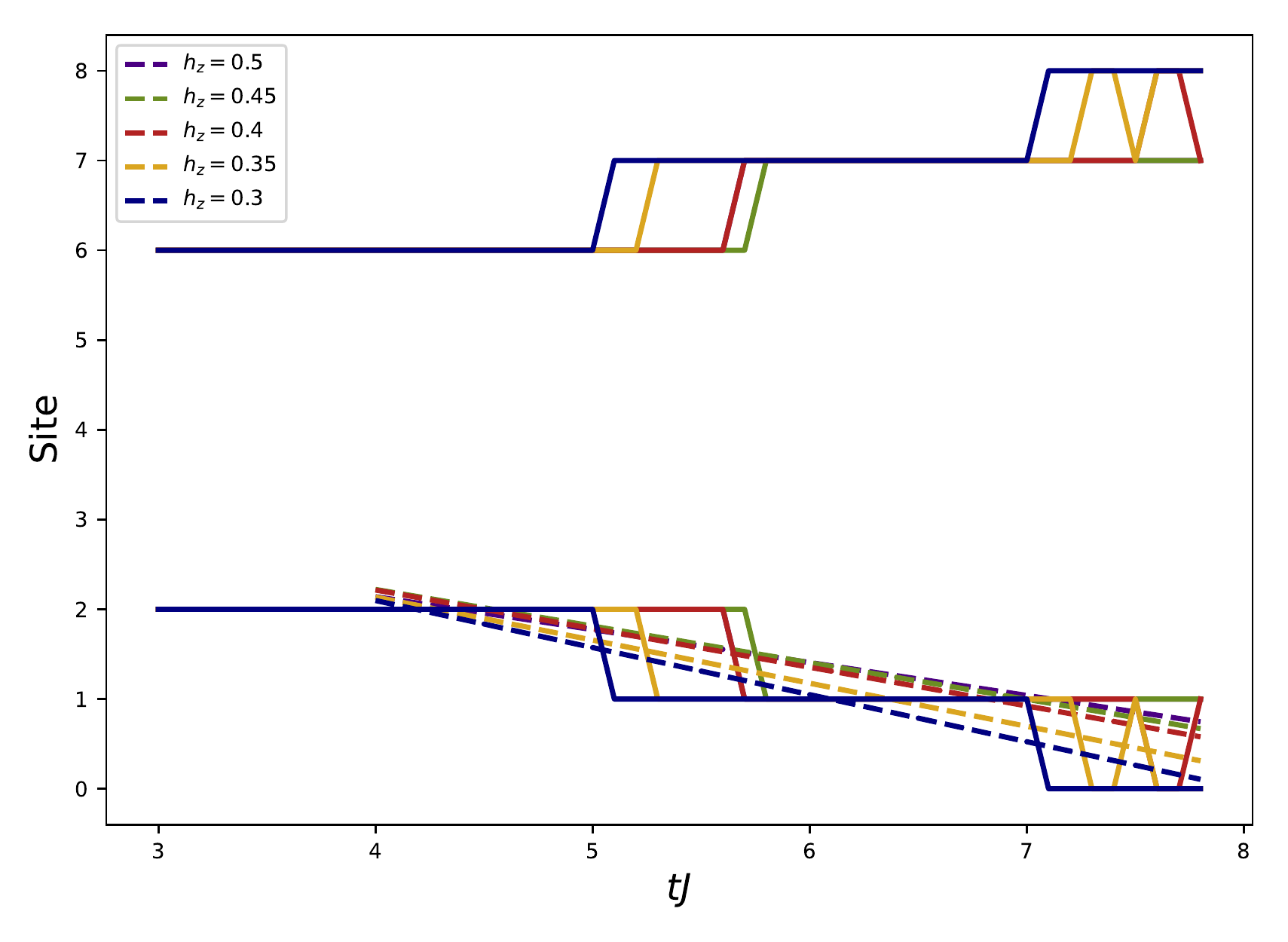}
    \caption{\textbf{Light cones produced by saturating $\Delta_i^{zz}$ data}. Data saturated at $S=0.21$ for $\Delta_{i}^{zz}$ after a global quantum quench to the TFIM with a longitudinal field on the state $\ket{\frac{L-1}{2},1}$. In all presented data $h_x=0.5$, and $L=9$. Clear light cones are formed with varying velocities that depend on the strength of the longitudinal field.}
    \label{Fig7}
\end{figure}

It turns out that the meson velocities are much easier to obtain because the meson velocities themselves are observed at a larger scale. Data was collected up to $t=8J$ using seven trotter steps. An initial time of $t=4J$, roughly the time at which mesons form, was used as the starting point of the light cone to measure the meson velocities. The light cones produced for a given saturation level, $S = 0.21$, and the corresponding velocities are shown in Fig.\ref{Fig7}. Again, the same range of saturation levels was used for each value of $h_z$ to obtain an average velocity.

\newpage
\paragraph*{Validity of the two kink projection.---} In all of the results presented for $\Delta_i^{zz}$, probability maps and local magnetisation, we employed the error mitigation of post selecting the data that lie within the two kink subspace. In this section, we discuss the validity of the method. To quantify to what extent the two kink subspace dictates the motion of kinks one can look at $\phi(t)$ defined as

\begin{equation}
    \phi(t) = \sum_i |\bra{i}\ket{\psi(t)}|^2.
\end{equation}

\begin{figure}[H]
    \centering
    \includegraphics[width=1 \textwidth]{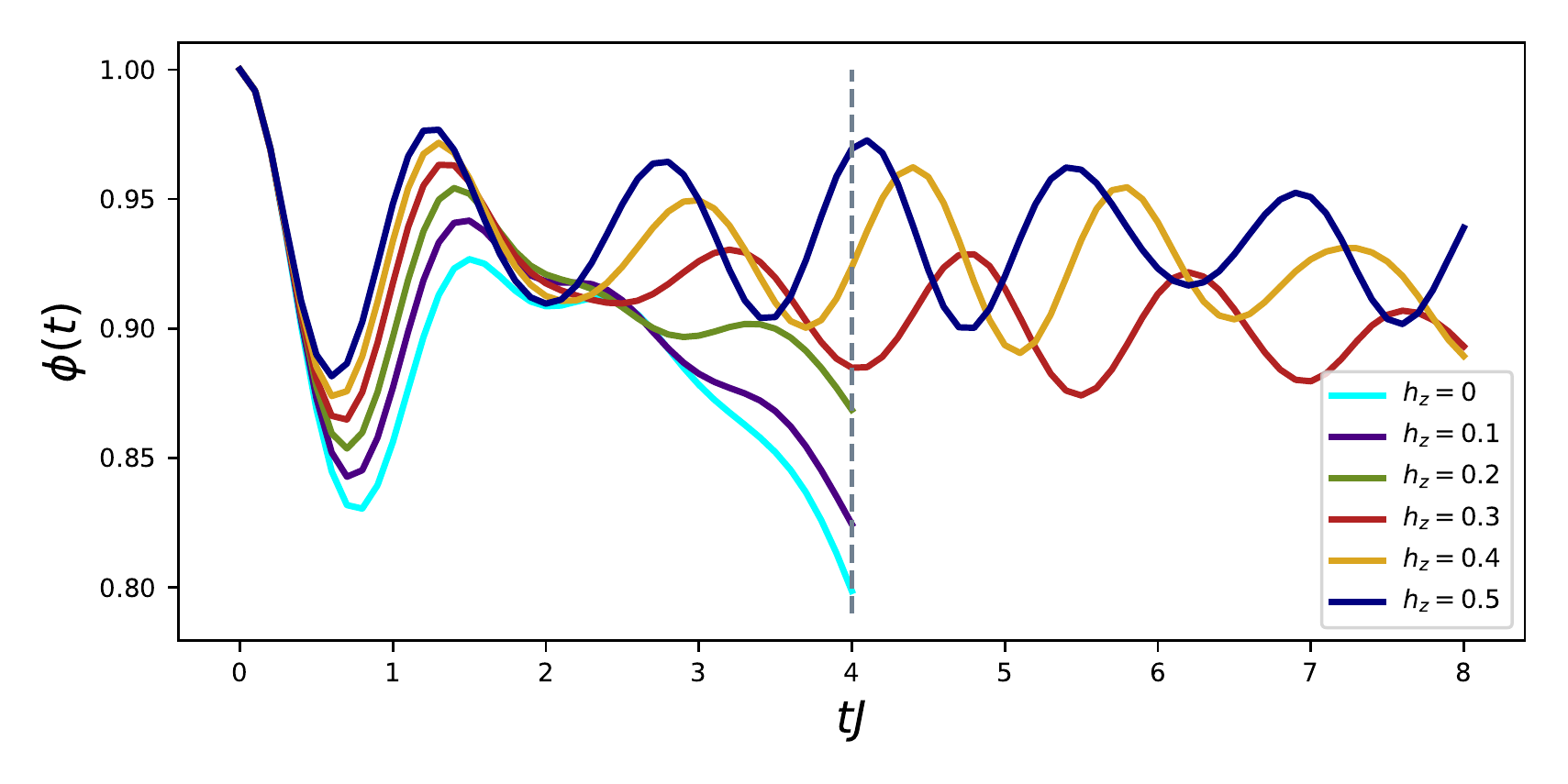}
    \caption{\textbf{A measure of the contribution of the two kink subspace}. $\phi(t)$ after a global quantum quench to the TFIM with a longitudinal field on the state $\ket{\frac{L-1}{2},1}$. Here $h_x=0.5$ and $L=9$. As only initial velocities were calculated for $h_z < 0.3$, $\phi(t)$ is presented in the range $tJ < 4$ for these longitudinal field strengths, this is highlighted by the horizontal dashed line. Even in the free kink case, $h_z=0$, the majority of the dynamics, $>80\%$, is within the two kink subspace for the times used to compute the velocities in Fig.\ref{Fig1} (b).}
    \label{Fig8}
\end{figure}

\begin{figure}[H]
    \centering
    \includegraphics[width=.5 \textwidth]{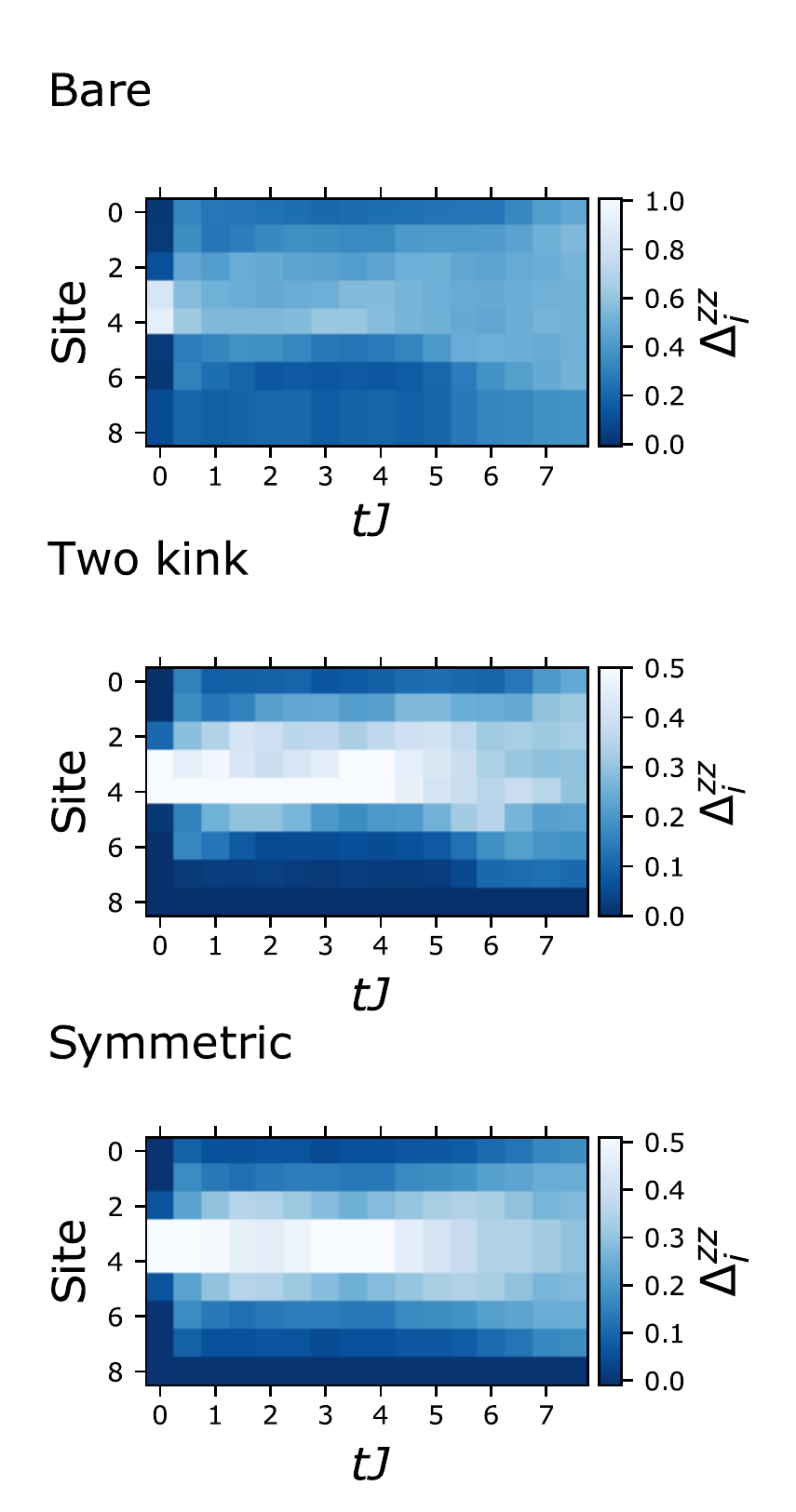}
    \caption{\textbf{A comparison of result collected from the IBM device with and without error mitigation}. The IBM results for$\Delta_{i}^{zz}$ presented in Fig.\ref{Fig3} with $h_x=0.5$, $h_z = 0.5$ and $L=9$. The bare result are shown first followed by the same data after post selection of the two kink subspace states, and finally the symmerised results. The background noise is considerably reduced by the two kink subspace projection and the light cones are restored by the process of forcing the inversion symmetry of the initial state. This allows the confinement physics to be observed and the gradients of light cones to be measured.}
    \label{Fig9}
\end{figure}

Here, $\ket{\psi(t)}$ is the the full many-body wave function of the spin chain at the time $t$ after the global quantum quench and $\big\{\ket{i}\big\}$ are the basis states of the two kink subspace. From this definition of $\psi$, $\phi(t)$ can be seen as a measure of the percentage of the full wave function, $\ket{\psi(t)}$, that remains in the two kink subspace. In Fig.\ref{Fig8} it is clear that the two kink subspace is responsible for $>80\%$ of the dynamics in the protocols used in this paper. Hence, by post selecting states measured in the two kink subspace the correct states will be selected up to reasonable error. Beyond this, in Fig.\ref{Fig9} a comparison of the results taken from the IBM device before and after error mitigation are presented. This highlights the effect of the subspace projection as well as enforcing the inversion symmetry of the initial state in order to project away background noise and extract the desired physical results. The power of this projection can be understood via the following argument. Let $\delta$ be the probability of measurement error for a qubit. For a two kink state there are just four possible erroneous spin flips that do not result in the measurement to be outside the subspace. Therefore,  to first order approximation the error that will not be mitigated is just $4\delta$ which does not scale with system size. In turn, the probability of error in simulations that can be mitigated via a projection into the two kink subspace is $(N-4)\delta$. At second order, the probability of two consecutive errors occurring that take the result out of and then back into the subspace is of order $\delta^2$. With a small $\delta$ this second order process is much less likely, and thus, this projection allows large error mitigation.

\newpage
\paragraph*{R\'enyi entropy measurements.---} Although the qualitative structure of R\'enyi entanglement entropy obtained on the IBM machine is correct, in order to obtain quantitative agreement with ED results a constant shift is needed. As the circuit depths used are too low for coherence times to be the source of this problem, the two possible causes are initialisation and measurement errors or gate errors. To understand this we performed two separate tests.

\begin{figure}
    \centering
    \includegraphics[width=1\textwidth]{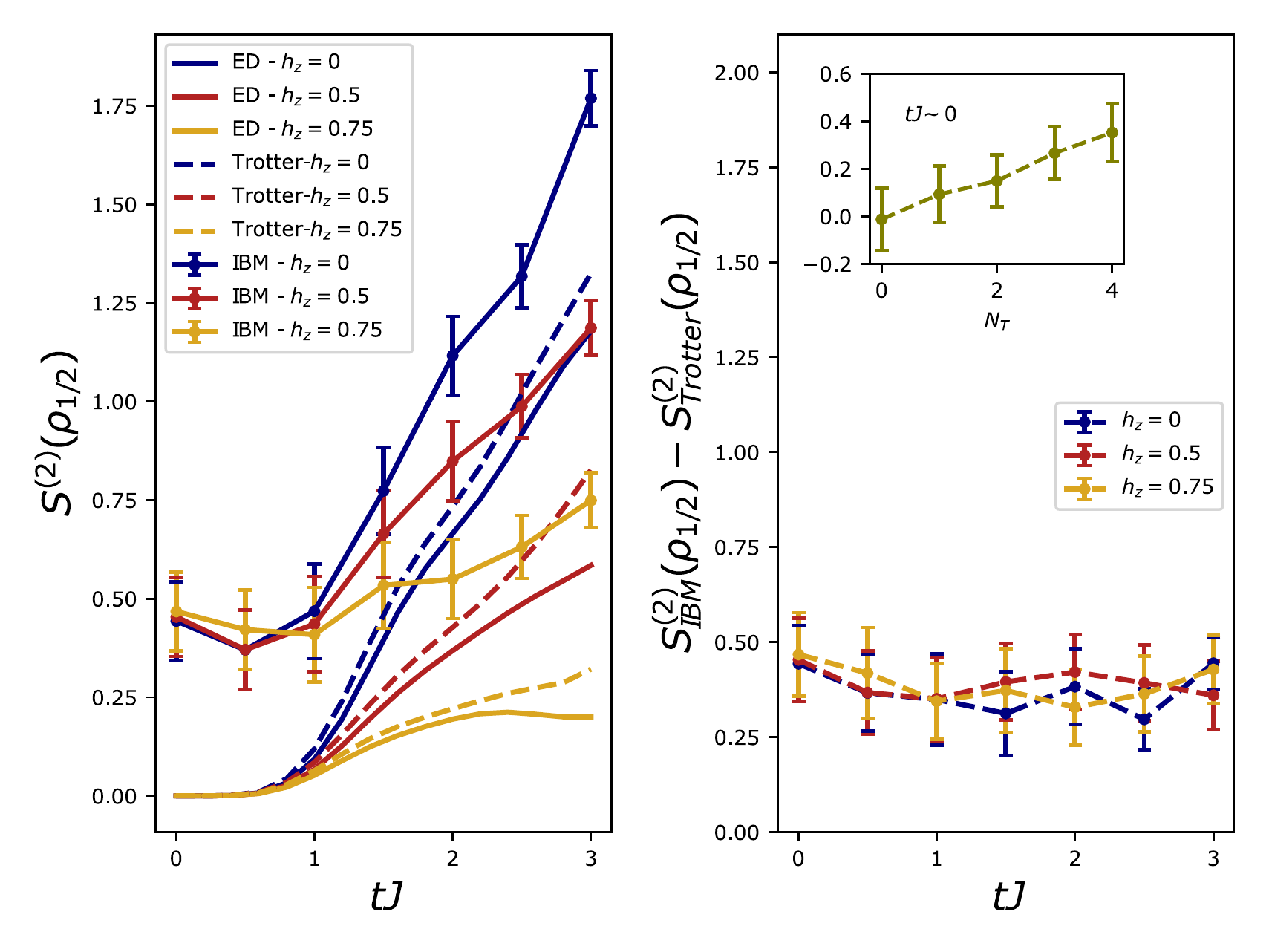}
    \caption{\textbf{Time evolution of the half chain R\'enyi entropy without shift mitigation}. The left panel shows the data for the second order R\'enyi entropy presented in Fig.\ref{Fig1}(c) with the constant shift from the IBM device left unresolved. The right panel shows explicitly this constant shift due to gate error with the inset plot showing how this shift depends on the number of trotter steps, $N_T$. This shift is time independent and grows with $N_T$, hence it is a results of implementation error on the IBM device.}
    \label{Fig10}
\end{figure}

Firstly, we ran a simple circuit in which we directly measure the entropy of the system initialised in the $\ket{\uparrow \uparrow \downarrow \downarrow \uparrow \uparrow}$ state with the expected result of zero. On the IBM machine this is equivalent to starting a run, initialising the state and adding the random single qubit gates on the half chain in consideration and measuring. As errors in single qubit gates are very low, any error in the entropy calculation can be assigned to initialisation and measurement error with a high degree of certainty. Using the built in measurement mitigation tool in Qiskit Ignis the entropy recorded is $\sim 0$ as seen in Fig.\ref{Fig10}. Thus, we conclude that this shift is not an effect of initialisation and measurement error. 

Secondly, we ran circuits with a varying number of trotter steps that only evolve the state to very short times, $tJ = 0.01$. This will also have an expected R\'enyi entropy of zero to the degree of accuracy which is obtainable by the IBM device. This protocol will include gate errors as well as initialisation and measurement errors. From Fig.\ref{Fig10} it can be seen that the error in performing gates on the IBM device has the effect of a constant entropy shift observed for all times measured on the device that grows with the number of trotter steps used. As this shift is not time dependent it is reasonable to assume it is due to errors in implementation only and can be safely removed to observe the true physics. In fact, a basic error model can be derived to account for this shift in random measurement protocols which will be presented elsewhere.

\begin{figure}
    \centering
    \includegraphics[width=0.5\textwidth]{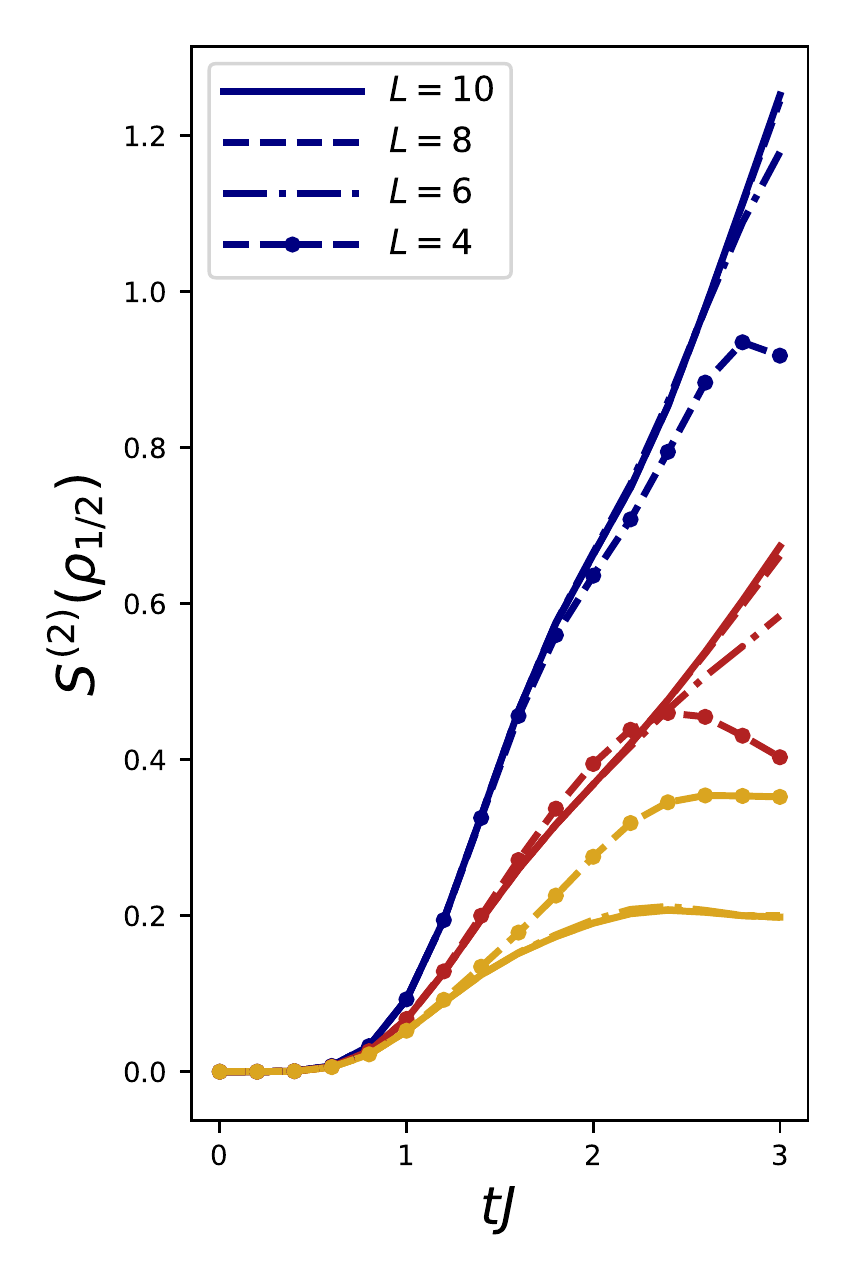}
    \caption{\textbf{Finite size scaling for the time evolution of the half chain R\'enyi entropy}. ED data for the second order R\'enyi entropy after a global quantum quench to the TFIM with varying longitudinal field strengths on the state $\ket{\frac{L}{2}-1,2}$ such that $L \in \{4,6,8,10\}$. Here, $h_x=0.5$. Blue results correspond to $h_z = 0$, red results correspond to $h_z = 0.5$ and gold results correspond to $h_z = 0.75$. Although finite size effects are present for a chain of length $L=6$ they only have a slight effect for times $tJ>2.5$ and are not large enough to hide the halting effect of entropy growth due to confinement.}
    \label{Fig11}
\end{figure}

As well as removing this shift, in order to obtain clear results for entropy, a system size of six sites was used. It is thus important to rule out finite size effect as the cause of the suppression in entropy. Fig.\ref{Fig11} shows the finite size scaling of R\'enyi entropy. For a system size of six times up to $tJ \sim 2.5$ are free of finite size effect. Although for longer times slight effects of a short chain are visible when considering longitudinal field strengths of $h_z = 0$ and $0.5$, they are small enough for the halting of entropy growth due to confinement to be visible. Furthermore, there are no finite size effects for a longitudinal field of strength $h_z = 0.75$. Hence, the suppression seen in the IBM device can be confidently assigned to confinement dynamics.

\end{document}